\documentclass[sn-basic, Numbered]{sn-jnl}

\usepackage{graphicx}%
\usepackage{multirow}%
\usepackage{amsmath,amssymb,amsfonts}%
\usepackage{amsthm}%

\usepackage{mathrsfs}%
\usepackage[title]{appendix}%
\usepackage{xcolor}%
\usepackage{textcomp}%
\usepackage{manyfoot}%
\usepackage{booktabs}%
\usepackage{algorithm}%
\usepackage{algorithmicx}%
\usepackage{algpseudocode}%
\usepackage{listings}%

\begin{document}

\title[Article Title]{Mathematical Methods for Assessing the Accuracy of Pre-Planned and Guided Surgical Osteotomies}

\author[1,2]{\fnm{George R.} \sur{Nahass}}\email{gnahas2@uic.edu}

\author[1]{\fnm{Nicolas} \sur{Kaplan}}\email{nkapla3@uic.edu}
\equalcont{These authors contributed equally to this work.}

\author[1]{\fnm{Isabel} \sur{Scharf}}\email{ischarf2@uic.edu}
\equalcont{These authors contributed equally to this work.}

\author[1]{\fnm{Devansh} \sur{Saini}}\email{dsaini2@uic.edu}
\equalcont{These authors contributed equally to this work.}

\author[1]{\fnm{Naji Bou} \sur{Zeid}}\email{nbouzeid@uic.edu}
\equalcont{These authors contributed equally to this work.}

\author[1]{\fnm{Sobhi} \sur{Kazmouz}}\email{skazmo2@uic.edu}

\author[1]{\fnm{Linping} \sur{Zhao}}\email{lpzhao99@uic.edu}

\author*[1,2]{\fnm{Lee} \sur{W.T. Alkureishi}}\email{lalk@uic.edu}

\affil*[1]{\orgdiv{Plastic Surgery}, \orgname{University of Illinois Chicago}, \orgaddress{\street{Paulina}, \city{Chicago}, \postcode{60612}, \state{IL}, \country{USA}}}

\affil[2]{\orgdiv{Biomedical Engineering}, \orgname{University of Illinois Chicago}, \orgaddress{\street{S Morgan}, \city{Chicago}, \postcode{60607}, \state{IL}, \country{USA}}}

\abstract{The fibula-free flap (FFF) is a valuable reconstructive technique in maxillofacial surgery; however, the assessment of osteotomy accuracy remains challenging. We devised two novel methodologies to compare planned and postoperative osteotomies in FFF reconstructions that minimized user input but would still generalize to other operations involving the analysis of osteotomies. Our approaches leverage basic mathematics to derive both quantitative and qualitative insights about the relationship of the postoperative osteotomy to the planned model. We have coined our methods 'analysis by a shared reference angle' and 'Euler angle analysis'. In addition to describing our algorithm and the clinical utility, we present a thorough validation of both methods. Code is available at https://github.com/monkeygobah/osteoplane.}

\keywords{plastic surgery, computer vision, point clouds, fibula free flap}

\maketitle

\section{Introduction}\label{sec1}

The fibula free flap (FFF) is a commonly used flap in reconstruction, serving as a versatile tool in the repair of bony defects. Compared to other prevalent bony flaps such as iliac crest and scapula bone grafts, FFF has demonstrated similar morbidity and improved stability postoperatively \cite{patel2019fibular, wilkman2017comparison, fujiki2013comparison, yilmaz2008comparison, politi2012iliac, brown2017mandibular}. The FFF also offers sufficient bone for free flaps, with bone, blood supply, and skin being relatively easy to harvest \cite{patel2019fibular}. As a result of this versatility, FFF has established itself as a popular technique for mandibular reconstruction \cite{patel2019fibular, brown2017mandibular, fujiki2013comparison}.

Advances in technology have been successfully integrated into the execution of the fibula free flap for mandible reconstruction. For instance, virtual surgical planning has been paired with computer-aided modeling and 3-dimensional (3D) printed cutting guides to preoperatively plan osteotomies for FFF \cite{logan2013exploratory, patel2019fibular, el2014evaluation, kaplan2023virtual}. Optimization of FFF utilizing the aforementioned techniques has led to more time-efficient operations with improved patient recovery and satisfaction \cite{patel2019fibular}. 

Despite the relative integration of these techniques into the practice of mandible reconstruction there is a relative scarcity in the literature as to how to specifically analyze the accuracy with which the plan has been executed. Further, there is significant diversity in the specific implementation of existing technologies into the surgical workflow: for instance, a surgeon may select to use virtual surgical planning, 3D printing of post-operative model or cutting guides, or a combination thereof. It is also worth noting that many older surgeons also elect to freehand the surgery without virtual planning or intraoperative guides.

General trends in the reporting of pre-surgical planning and post-operative outcomes include qualitative description of aesthetic outcomes, quantitative comparison between length of pre-vs-post fibular flap sections or relative positioning of the segments, as well as general operative metrics such as cost and time \cite{roser2010accuracy,antunez2021mandibular, guo2022design, bosc2016henri,  mottini2016new, wang2016virtual, seier2020virtual, ren2018virtual, lai2022mandible}. While these methodologies can provide a broad overview as to the success of the surgical plan and adjunct technology, there remains a degree of subjectivity and the potential for the surgeon to deviate from preoperative planning. This may be due to inaccurate patient scans, changes in resection secondary to continued neoplastic growth, and/or the need to adapt planning to prioritize patient facial aesthetics \cite{garcia2019craniosynostosis}. By examining the angle of the osteotomy planes, one can consider the extent to which the surgical planning-derived cutting guides have been accurately replicated \cite{guo2022design, bosc2016henri}. However, creating these planes can represent a significant challenge, requiring the acquisition of often-expensive software that can be complex to learn and implement. Furthermore, there can be significant variance in the reliability of the data depending on user selection of points in order to create an osteotomy plane, thus contributing to error. 

By implementing computational algorithms, one can circumvent some of these sources of error and decrease the effect of individual bias on outcome measures. We set out to devise a novel methodology for comparing planned and post-operative osteotomies in fibula-free flap (FFF) reconstructions without reliance on the overall position of the reconstruction, which is potentially affected by many more variables. We designed two computational solutions and bundled them in the same software package. ‘Analysis by a Shared Reference Plane’ (method 1) and ‘Analysis by Rotation’ (method 2) share some common steps, which can be seen in section \ref{data_preproc}.

\section{Methods}\label{methods}
\subsection{Study Design}\label{study_design} 
Both pre and post operative scans were analyzed to compare the accuracy of the surgical planning to the actual surgical outcome on the patient. Python 3.8.3 was used to write all code while the point cloud visualization and selection was performed using the Open3D python library \cite{zhou2018open3d}. The Plotly library was used to plot three-dimensional data. All experiments were performed on a 2021 M1 MacBook Air.

\begin{figure}[!t]
    \centering
    \includegraphics[width=8cm]{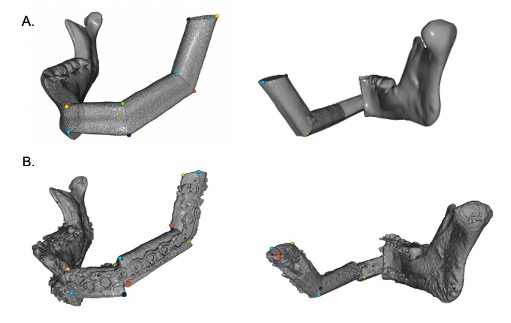}
    \caption{Planned FFF reconstruction point cloud with points selected defining the planes of interest. A) is the planned model with 12 points selected, B) is the post operative model with 18 points selected. These points are used to define the planes according to equation \ref{first_set}.}
    \label{fig:point_select}
\end{figure}

A 59 year old male presenting with squamous cell carcinoma of the buccal mucosa with invasion of the mandible required a corrective left FFF. A preoperative CT scan of the head and neck was collected, and the DICOM file from this scan was subsequently imported into “ImmersiveRecon” (ImmersiveTouch, Chicago IL) for the surgeon to plan using virtual reality. The subsequent plan was used to produce custom 3D-printed cutting guides for intraoperative use; an STL file of the planned mandible repair was uploaded to our algorithm. Post-op CT scans of the head and neck were taken, and those DICOM files were imported to Mimics (Materialise NV, BE) for manual cleaning of any artifact from CT scanning. After cleaning, the mandible was segmented and exported as an STL file for downstream analysis.

\subsection{Data Preprocessing}\label{data_preproc} 
In order to define the points that make the plane, the user selects 3 points on each model (plan and post-operative) to define the segments that will later be analyzed. As the planned model is the idealized surgical outcome, for a 3-segment FFF reconstruction, only 4 planes are needed (\textbf{Figure \ref{fig:point_select}}A), however, for the postoperative model, 18 points are needed on a 3 segment reconstruction to define the individual segments (\textbf{Figure \ref{fig:point_select}}B). STL files are input and then visualized as point clouds of $1x10^6$ density to allow for point selection (Open3D Python Package). The anatomical landmarks must be selected in order of the anterior border, lateral border, and medial border of the fibula, respectively, to allow for accuracy in the following steps. Plan and post-operative models with selected planes visualized as triangles are shown in \textbf{Appendix Figure \ref{fig:point_planes_prox}}.

\begin{figure}[!t]
    \centering
    \includegraphics[width=10cm]{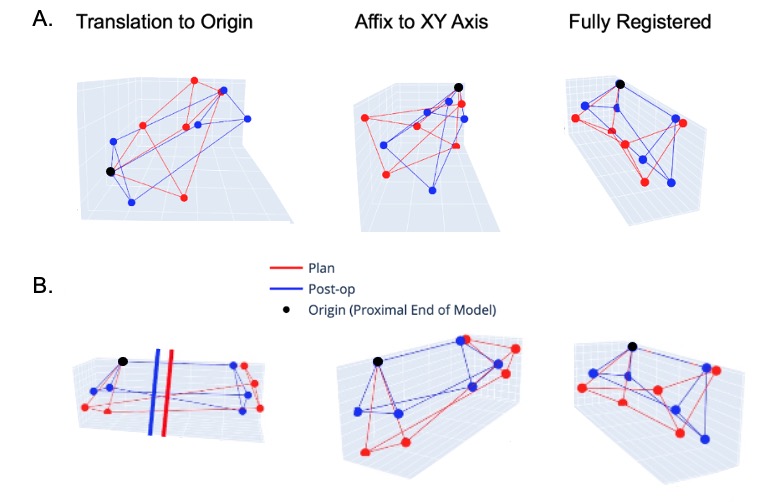}
    \caption{A) Demonstrate the steps taken to register a single segment of pre and post-operative models. This same procedure is performed for all segments to ensure accurate downstream analysis. B) Multiple views of a fully registered segment across two models. Note the longitudinal line from the origin for both plan and postoperative segments overlaps; this represents the AS border of the model, and the overlap confirms a successful registration of the segments.}
    \label{fig:registration}
\end{figure}

Following the selection of points, we continue forward in the analysis only using the selected points as they should convey all the information we need to determine the relationship of the osteotomies between postoperative and planned. Another preparation step is needed, which is the registration between the segments. To allow for accurate registration between models, each segment is isolated and analyzed individually using the initially selected points. We use the anterior border of the fibula to permit registration of the segments. 

The registration strategy consists of 3 parts: 
\begin{enumerate}
    \item  Translation of the whole segment to the origin 
    \item Alignment of the anterosuperior border (AS) of the planned and post-operative planes to the $XY$ plane of the grid followed by rotation of all other points 
    \item Alignment of the AS border of the planned and post-operative planes to the $y$ axis followed by a rotation of all other points. 
\end{enumerate}    
We refer to proximal and distal relative to the symphysis of the mandible where proximal is the plane closest to the symphysis on any given segment (\textbf{see Appendix Figure \ref{fig:point_planes_prox}}). Our registration strategy results in the placement of the proximal point of the AS border of the planned and post-operative models at the origin and the distal point of the AS border at some length d along the $y$-axis.

Translation of the models to the origin is done by subtracting the coordinates of the anterior border point of the proximal plane of both the planned and post-operative model from all other points in the respective segment. Alignment of the AS border of the planned and post-operative planes to the $XY$ plane and rotation of all points of the segment is accomplished according to Equation \ref{seg_rot}, and the final alignment to the $y$ axis is accomplished via Equation \ref{affix} where $x,y,z$ represent the coordinates of the anterior border of the distal plane. Visualization of all of these steps can be seen for a single segment as well as a panorama view of a fully registered segment can be seen in \textbf{Figure \ref{fig:registration}}.

\begin{align}
\begin{minipage}{.45\linewidth}
\begin{equation}
\begin{aligned}
\theta &= -\arctan{\frac{z}{y}} \\
x' &= x \\
y' &= y\cos{\theta} - z \sin{\theta} \\
z' &= y \sin{\theta} + z \cos{\theta}
\end{aligned}
\label{seg_rot}
\end{equation}
\end{minipage}
\quad
\begin{minipage}{.45\linewidth}
\begin{equation}
\begin{aligned}
\theta &= -\arctan{\frac{x}{y}} \\
x' &= x \cos{\theta} - y \sin{\theta} \\
y' &= y\cos{\theta} - z \sin{\theta} \\
z' &= z
\end{aligned}
\label{affix}
\end{equation}
\end{minipage}
\end{align}

\section{Results}\label{sec2}
\subsection{Method 1: Analysis by a Shared Reference Plane}\label{method_1} 

As the AS border of both the planned and postoperative models are now aligned to the $y$ axis (on a per-segment basis), this forces the $XZ$ plane to be perpendicular to the longitudinal aspect of the segment for both the planned and postoperative model. The plane defining the proximal and distal ends of the segment to be analyzed are defined based on point selections from the user according to Equation \ref{first_set} where $\mathbf{a, b, c}$ represent the normal vector. As the $XZ$ plane is perpendicular to both the planned and postoperative model, we define it herein as $\mathbf{v}_{\text{ref}}$. The difference of the planes defining the proximal and distal segments and the $XZ$ plane can now be computed for both planned and postoperative models. This is accomplished via Equation \ref{second_set} where $d_{model}$ represents the $ y$ length of the AS border obtained from user point selections.

\begin{align}
\begin{minipage}{.45\linewidth}
\begin{equation}
\begin{aligned}
p_1 &= (x_1, y_1, z_1) \\
p_2 &= (x_2, y_2, z_2) \\
p_3 &= (x_3, y_3, z_3) \\
\mathbf{v}_1 &= p_3 - p_1 \\
\mathbf{v}_2 &= p_2 - p_1 \\
a, b, c &= \frac{\mathbf{v}_1 \times \mathbf{v}_2}{\|\mathbf{v}_1\| \|\mathbf{v}_2\|}
\end{aligned}
\label{first_set}
\end{equation}
\end{minipage}
\quad
\begin{minipage}{.45\linewidth}
\begin{equation}
\begin{aligned}
\text{Ref}_{\text{equation}} &: Ax + Cz = \frac{d_{\text{model}}}{2} \\
\mathbf{v}_{\text{ref}} &= (a_2, b_2, c_2) = \left(0, \frac{d_{\text{model}}}{2}, 0\right) \\
\mathbf{v}_{\text{model}} &= (a_1, b_1, c_1) \\
q &= \frac{\mathbf{v}_{\text{ref}} \cdot \mathbf{v}_{\text{model}}}{\|\mathbf{v}_{\text{ref}}\| \|\mathbf{v}_{\text{model}}\|} \\
\theta &= \arccos{q}
\end{aligned}
\label{second_set}
\end{equation}
\end{minipage}
\end{align}

\begin{figure}[!t]
    \centering
    \includegraphics[width=9cm]{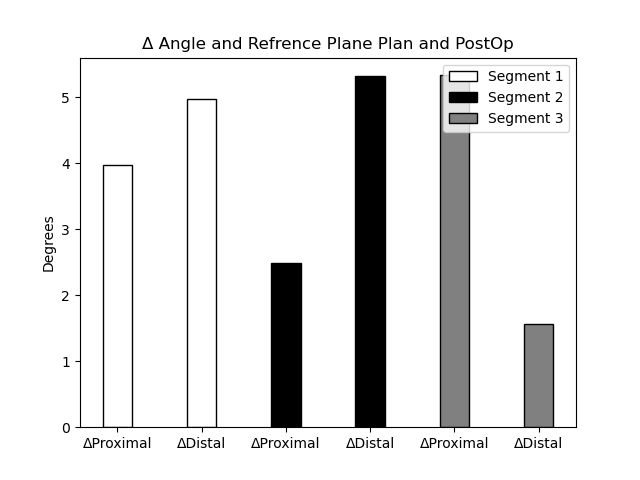}
    \caption{Results of a single procedure analyzed using Method 1 described in section \ref{method_1}. $\Delta$ represents the difference between the angle of planned and postoperative planes with their shared reference plane ($XZ$ plane).}
    \label{fig:method_1_res}
\end{figure}

Using this framework, the angle between the proximal and distal planes relative to the $XZ$ plane can be computed for every segment on the planned and postoperative models. In order to assess the accuracy of the osteotomy, the absolute value of the difference between the angles on their respective segments and position can be calculated, where values closer to 0 indicate a lower deviation between the planned and postoperative osteotomy. This system provides minimal information about the geometric deviance from postoperative to planned model, but it does provide a single value that can be quickly understood to help the evaluator (who may be a surgeon or engineer) understand which osteotomies have a higher deviance from the planned model. An example of the output is shown in \textbf{Figure \ref{fig:method_1_res}}.

\subsection{Method 2: Analysis by Rotation Using Euler Angles
}\label{method_2} 

While method 1 presented above provides a rapid method to compare the relationships of related planes between pre and post-operative models, it fails to describe the 3-dimensional relationship between the osteotomies due to the use of a single angular output. To attempt to address this, we designed a methodology that uses the same planes calculated via 3, but instead computes the Euler angles required to rotate the postoperative plane to become parallel to the planned plane for each proximal and distal osteotomy plane per segment. Initially, the axis and angle of rotation for the postoperative plane relative to the preoperative plane must be computed. This is done via equation \ref{euler_planes} where $\mathbf{v}_{\text{plan}}$ and $\mathbf{v}_{\text{postop}}$ represent the normal vectors of the planes. Using the unit vectors, the axis of rotation is identified by taking the normalized cross-product. The angle of rotation is represented by $\Theta$ in Equation \ref{euler_planes}.

\begin{equation}
\begin{aligned}
\mathbf{v}_{\text{plan}} &= \frac{(a_{\text{plan}}, b_{\text{plan}}, c_{\text{plan}})}{\|\mathbf{a}_{\text{plan}}, \mathbf{b}_{\text{plan}}, \mathbf{c}_{\text{plan}}\|} \\
\mathbf{v}_{\text{postop}} &= \frac{(a_{\text{postop}}, b_{\text{postop}}, c_{\text{postop}})}{\|\mathbf{a}_{\text{postop}}, \mathbf{b}_{\text{postop}}, \mathbf{c}_{\text{postop}}\|} \\
\mathbf{axis} &= \frac{\mathbf{v}_{\text{plan}} \times \mathbf{v}_{\text{postop}}}{\|\mathbf{v}_{\text{plan}} \times \mathbf{v}_{\text{postop}}\|} \\
\Theta &= \arccos(\mathbf{v}_{\text{plan}} \cdot \mathbf{v}_{\text{postop}})
\end{aligned}
\label{euler_planes}
\end{equation}

We then employ the Rodriguez Formula in equation \ref{rodrigues} to derive a rotation matrix from the axis and angle of rotation. We first create a skew matrix using the axis of rotation and compute the rotation matrix. The euclidean norm of the projection of the third column vector of the rotation matrix onto the $XY$ plane is calculated in order to check for gimbal lock which would result in a loss of a degree of freedom. The presence of absence of gimbal lock is checked internally within the algorithm by assessing whether or not $||R_3||xy$ is less than 1x10-6. If gimbal lock is present, the psi, theta and phi angles are calculated according to equation \ref{gimbal} instead of \ref{rodrigues}.  $\Phi$, $\theta$, and $\Psi$ can be interpreted as the rotations that happen in the z, y, and x dimensions, respectively, to rotate the post-operative plane to become parallel to the planned plane. Planes from the plan, post-operative, and rotated postoperative planes from a subset of segments are displayed in \textbf{Appendix Figure \ref{fig:euler_rotated_vis}}.

\begin{align}
\begin{minipage}{.45\linewidth}
\begin{equation}
\begin{aligned}
\mathbf{S} &= \begin{pmatrix}
0 & -\text{axis}_z & \text{axis}_y \\
\text{axis}_z & 0 & -\text{axis}_x \\
-\text{axis}_y & \text{axis}_x & 0
\end{pmatrix} \\
\mathbf{R} &= \mathbf{I} + \sin(\Theta) \cdot \mathbf{S} + (1 - \cos(\Theta)) \cdot \mathbf{S}^2 \\
\|\mathbf{R}_3\|_{xy} &= \sqrt{R_{1,3}^2 + R_{2,1}^2} \\
\Phi &= \arctan\left(\frac{R_{3,2}}{R_{3,3}}\right) \\
\theta &= \arctan\left(\frac{-R_{3,1}}{\|\mathbf{R}_3\|_{xy}}\right) \\
\Psi &= \arctan\left(\frac{R_{2,1}}{R_{1,1}}\right)
\end{aligned}
\label{rodrigues}
\end{equation}
\end{minipage}
\quad
\begin{minipage}{.45\linewidth}
\begin{equation}
\begin{aligned}
\Phi &= \arctan\left(\frac{-R_{2,3}}{R_{2,2}}\right) \\
\theta &= \arctan\left(\frac{-R_{3,1}}{\|\mathbf{R}_3\|_{xy}}\right) \\
\Psi &= 0
\end{aligned}
\label{gimbal}
\end{equation}
\end{minipage}
\end{align}

\begin{figure}[!t]
    \centering
    \includegraphics[width=\textwidth]{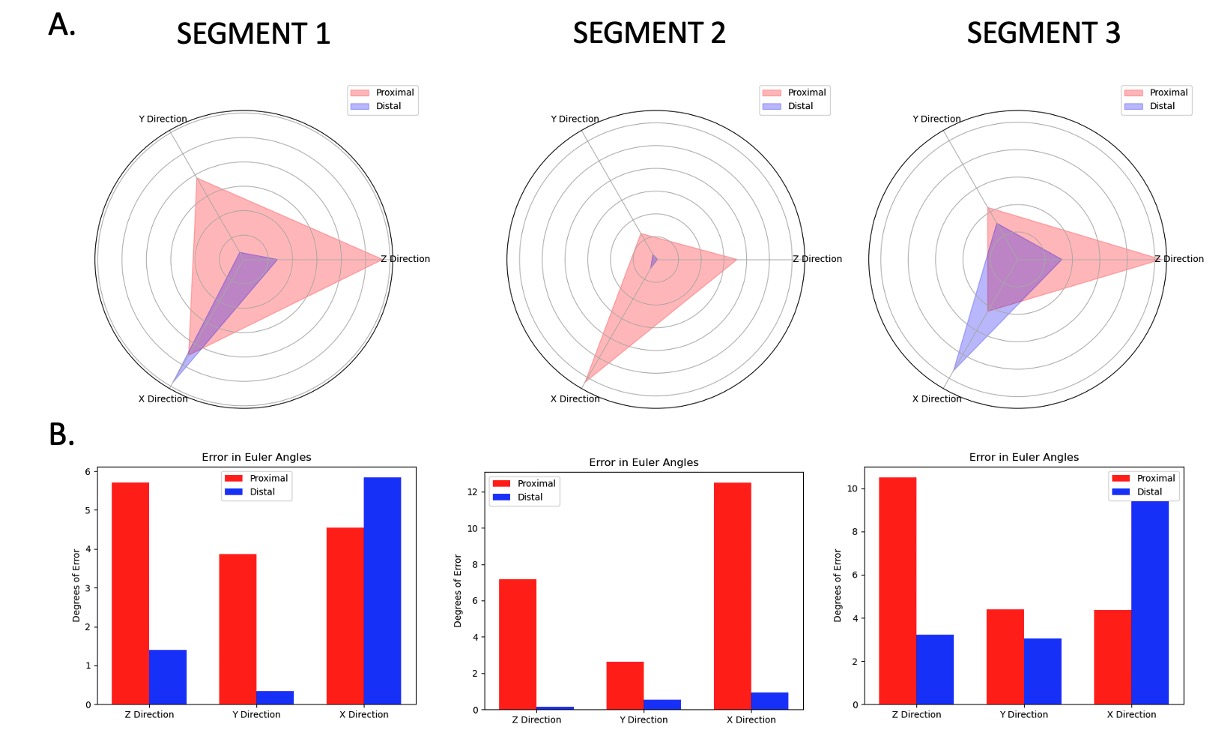}
    \caption{Output of Method 2 (Euler angle analysis) for a 3-segment FFF reconstruction. A) Radar and B) bar plots show both qualitative and quantitative aspects of the deviation of the surgery from planned to post-operative model on both planes for all 3 segments.}
    \label{fig:method_2_res}
\end{figure}

The benefit of using Method 2 is that it provides the surgeon or engineer with 3 discrete angles for the proximal and distal ends of each segment in the FFF reconstruction, which represent information about the locality of error. As opposed to Method 1, which only provides information about the magnitude of the error, Method 2 provides information about spatial trends in which deviance from the plan is intraoperatively introduced. A single FFF reconstruction was analyzed using Method 2, and the results are displayed in \textbf{Figure \ref{fig:method_2_res}}. The radar plot demonstrates the magnitude of the Euler angles for proximal and distal planes and shows quickly at a glance the axis in which the planes were most prominently rotated. The bar graphs display the true values for both proximal and distal planes in all three segments.

\subsection{Intraobserver Repeatability}\label{repeat} 
\begin{figure}[!t]
    \centering
    \includegraphics[width=10cm]{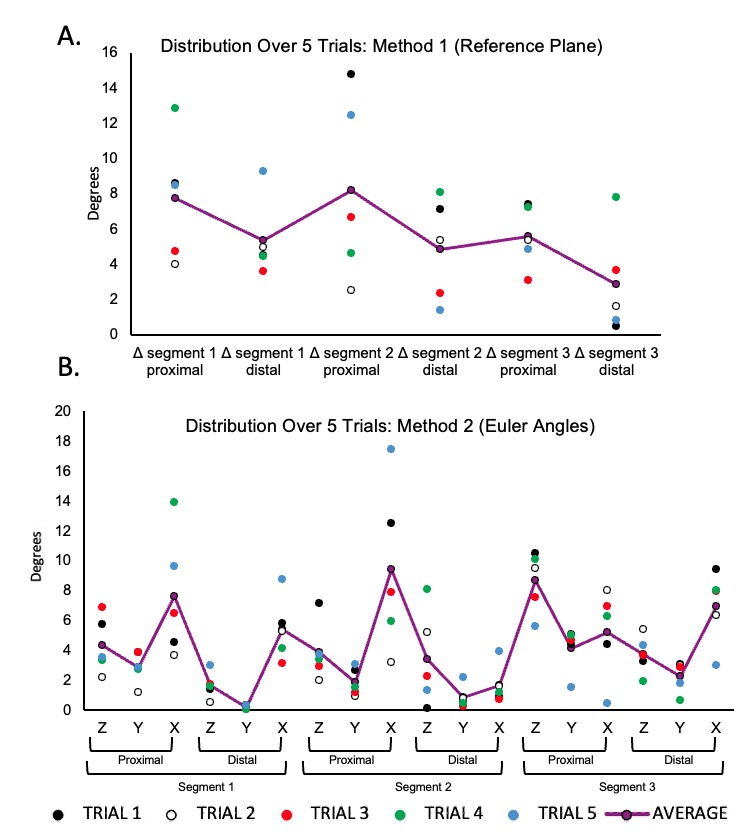}
    \caption{Average and standard deviation for both Method 1 (A) and Method 2 (B) displayed for each segment in a 3 segment FFF reconstruction over 5 independent trials by a single user. Purple bar represents the average error.}
    \label{fig:error}
\end{figure}

In an intraobserver repeatability test we analyzed the results of method 1 and method 2 for each segment in a FFF reconstruction over 5 trials (\textbf{Figure \ref{fig:error}}). 

\section{Discussion}\label{disc}

Evaluation of the accuracy of postoperative results compared to preoperative plans is one that has been investigated in the orthopedic literature.  Ferner et al (2023) examined femoral and tibial torsion by assessing the angle of CT scans at different slices to determine the agreement between the intended and postoperative values. However, they also noted a scarcity of the literature in the analysis of postoperative versus planned outcomes. Notably, their analysis was primarily based on cross-sectional slices of the CT scans.

By visualizing the free flap positioning in its entirety as we do in our methods, the surgeon can better understand the deviation of the plane from the plan and its clinical implications for future operations. An alternative approach to preoperative planning is to examine the difference between 3D position of the plan versus postoperative outcome; however, this does not specifically identify positional variables that represent surgical intervention such as the angulation of the osteotomy plane. Our methodology allows the viewer to examine the model in full and to identify specific deviations that might be subverted by intraoperative clinical adjustments. 

Here we present two novel mathematical methods for the analysis of intraoperative osteotomies made during FFF repairs. Method 1 provides high-level information about the deviation of the intraoperative osteotomy from the planned model by establishing a shared reference plane between both models. Method 2 calculates the Euler angles required to rotate the postoperative osteotomy to become parallel to the planned model, introducing a directional element into the analysis. Both strategies use our novel registration scheme (described in section \ref{data_preproc}), which takes advantage of the anatomical properties of the fibula. The outputs of our algorithm can be analyzed by the surgeon for postoperative analysis: by examining trends across multiple cases, the surgeon may elucidate information about consistent deviations in directionality from the pre-operative plan and attempt to mediate digression from the planned model. By utilizing the mathematical description of user-defined planes from FFF repairs, our method allows users to evaluate the accuracy of individual osteotomy segments without relying on the overall position of the reconstruction. This is highly advantageous due to the numerous factors that may influence the outcome of FFF reconstructions, such as intraoperative change and numerous individuals working on the same operation.

\section{Conclusion}\label{conc}
When operated by a trained user, both of our methods have a low error and are reproducible. However certain segments in both methods, such as the proximal plane on segment 1 and 2, do display a wider range of calculated angles following multiple trials (\textbf{Figure \ref{fig:error}}). This variability stems from the quality of the postoperative CT scan, and the challenges in defining the plane in regions where the distal plane of one segment is occluded by the proximal plane of another. Hard to identify anatomical features of the mandible post operative will always introduce a degree of bias into the analysis when the plane equations are derived from user selected points. As such, having high quality scans (both pre and postoperative) as input models is a requirement to achieve optimal results. Given this innate bias introduced by user input in the initial point selection, the most accurate results using our method are obtained by taking the average of multiple trials and making sure an experienced operator selects the borders of the plane following reconstruction. Additionally, for high fidelity registration of the planned and post operative model it is imperative that the anterior border of each segment is selected first when defining the planes. Manual registration of point clouds does allow for a user to exert control over the selection of the planes defining the segments, but it does also allow for variation between runs as it is non-deterministic.



To mitigate bias introduced from user point selection, deep learning and automatic point cloud registration present an attractive future direction for the evaluation of FFF repairs. The literature already reports successful usage of deep learning for analysis of human craniofacial anatomy to generate reference models and predict a patient's need for surgery \cite{xiao2021estimating, shin2021deep}. A major limitation of deep learning, however, is the availability of training data. While our algorithm can of course not generate novel CT scan data, features such as the angles between planned and postoperative models can be leveraged to train future models to predict such measures.

\bibliography{osetoplane-article}

\newpage 

\begin{appendices}

\section{Supplemental Figures}\label{secA1}
\begin{figure}[!b]
    \centering
    \includegraphics[width=9cm]{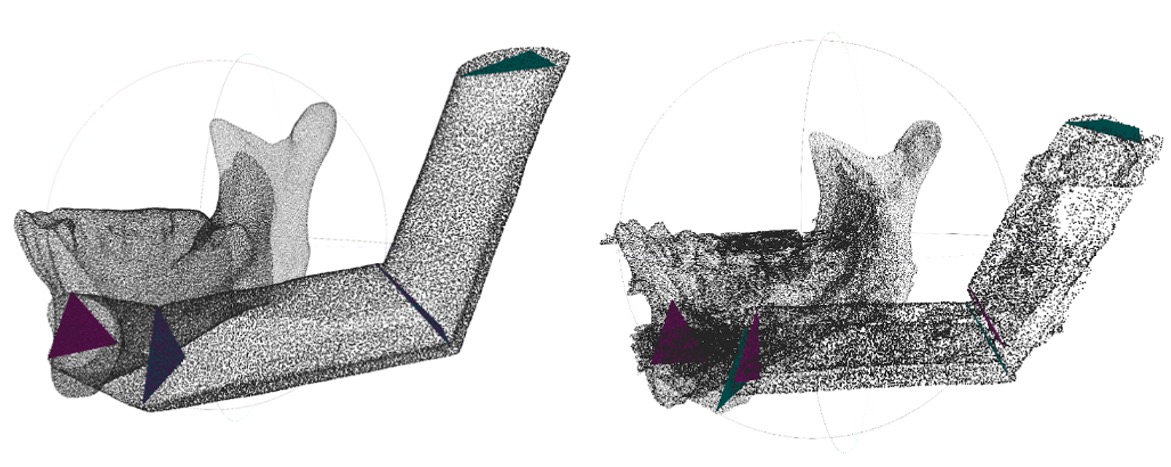}
    \caption{Point clouds of planned (left) and post operative (right) models showing the proximal (purple) and distal (green) user defined planes.}
    \label{fig:point_planes_prox}
\end{figure}

\begin{figure}[!b]
    \centering
    \includegraphics[width=9cm]{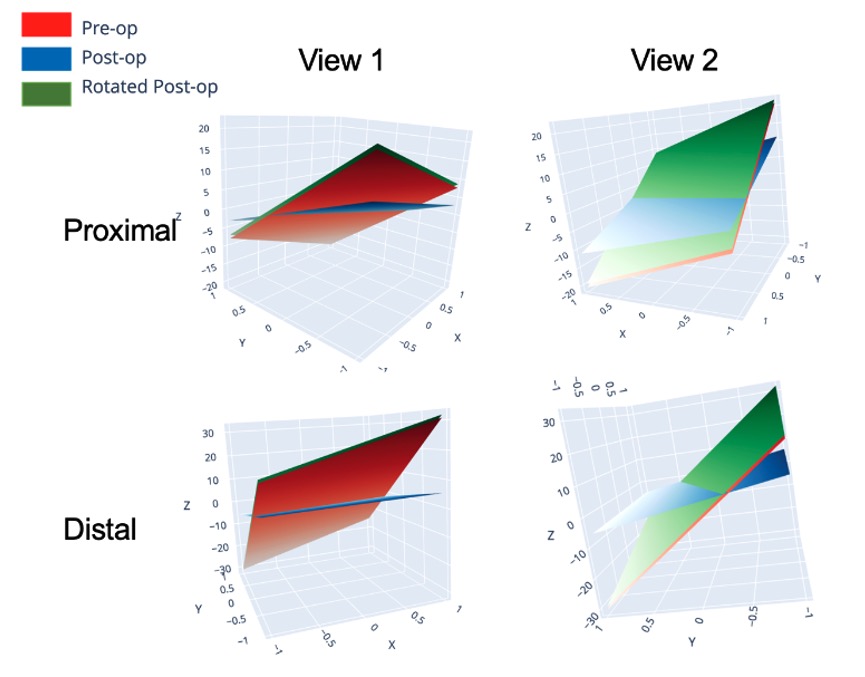}
    \caption{Results of a single segment for a single model demonstrating the calculated Euler angles result in the desired system of rotations to make the postoperative plane (blue) parallel to the planned model (red). The rotated post operative plane can be seen in green.}
    \label{fig:euler_rotated_vis}
\end{figure}

\end{appendices}

\newpage 













\end{document}